\documentstyle[aps,pra,epsf]{revtex}
\begin{document}
%\draft\twocolumn
\newcommand{\ket}[1]{\left | \, #1 \right \rangle}
\newcommand{\bra}[1]{\left \langle \, #1 \right | }
\newcommand{\be}{\begin{equation}}
\newcommand{\ee}{\end{equation}}
\newcommand{\bea}{\begin{eqnarray}}
\newcommand{\eea}{\end{eqnarray}}
\newcommand{\bma}{\begin{mathletters}}
\newcommand{\ema}{\end{mathletters}}
\newcommand{\ratio}{\frac{\Delta} {\Omega}}

\title{Distributed Quantum Computation\\
over Noisy Channels}
\author{{J. I. Cirac$^1$, A. K. Ekert$^2$, S. F. Huelga$^3$\cite{OVI} and C. Macchiavello$^4$}}
\address{
$^1$Institut f\"{u}r Theoretische Physik, Universit\"{a}t Innsbruck,
Technikerstrasse 25, A-6020, Innsbruck, Austria.\\
$^2$Department of Physics, Clarendon Laboratory, University of Oxford,
Oxford OX1 3PU, U. K.\\
$^3$Optics Section, The Blackett Laboratory, Imperial College, London
SW7 2BZ, U. K.\\
$^4$Dipartimento di Fisica ``A. Volta'' and INFM - Unit\`a di Pavia, 
Via Bassi 6, 27100 Pavia, Italy.}

\maketitle
\begin{abstract}
We analyse the use of entangled states to perform quantum computations
non locally among distant nodes in a quantum network. The complexity
associated with the generation of multiparticle entangled states is
quantified in terms of the concept of global cost. This parameter allows
us to compare the use of physical resources in different schemes. We
show that for ideal channels and for a sufficiently large number of
nodes, the use of maximally entangled states is advantageous over
uncorrelated ones. For noisy channels, one has to use
entanglement purification procedures in order to create entangled states
of high fidelity. We show that under certain circumstances a
quantum network supplied with a maximally entangled input still yields a
smaller global cost, provided that $n$ belongs to a given interval $n\in
[n_{min},n_{max}]$. The values of $n_{min}$ and $n_{max}$ crucially depend on
the purification protocols used to establish the $n$-- processor
entangled states, as well as on the presence of decoherence processes 
during the computation. The phase estimation problem has been used to 
illustrate this fact.
\end{abstract}

% ==========================================================================
\section{Introduction}

Consider a quantum computation which can be divided into subroutines so
that each subroutine can be run on a separate quantum processor. The
processors may be placed at different locations/nodes of a computational
network and may exchange data with a selected central processor
\cite{grover:97}. Each processor operates on a partial input which has a
fixed size. The partial inputs may be independent of each other,
correlated or even entangled. When the computation is finished, the
central processor, after collecting partial outputs from the other
processors, stores the global output. This type of distributed
computation may be repeated several times to yield a desired result and
as such it features frequently in quantum parameter estimation
procedures e.g. the phase estimation in frequency standards
\cite{colorado,fs}. In some computational tasks, e.g. estimating a given
parameter with a prescribed precision, the number of repetitions depends
on the form of the input state --- some entangled states require less
repetitions than uncorrelated inputs. In the case of correlated input
states, we have to pre-compute the input state for each run of the
computation and this involves an additional use of physical resources.
Are we still better off when the complexity of the pre-computation is
included? How shall we include and compare the use of different physical
resources?

In this paper we quantify this complexity by introducing the notion of
cost of physical operations, such as the cost of establishing an
entangled pair over a channel, the cost of transmitting one classical
bit between components, the cost of running a quantum processor, etc.
and discuss the performance of the distributed quantum computation when
the inter-processor quantum communication is prone to errors, i.e. when
the quantum channels among the processors are noisy. 

This paper has been organized as follows. 
In section II we introduce the notion of cost in distributed quantum
computation and show that if
the ratio between the number of repetitions for the entangled and
separate inputs decreases fast enough with the size of the network then,
above some critical size, the computation can be made cheaper using
entangled inputs. This behaviour is illustrated explicitly
in the case of estimation of a small phase shift for qubits, which was 
considered in Ref. \cite{grover:97}.
In section III we introduce the phase estimation problem for disentangled and
maximally entangled states and show how to compare the two scenarios.
In section IV we analyse the ideal case of noiseless channel and error-free
computations. Noise along the channels linking the nodes of the network is 
taken into account in section V and the use of different purification schemes
is considered. We show that
for certain purification protocols it is advantageous to use entangled
states, whereas for others the cost of the precomputation is not
off--set by the subsequent reduction in the number of repetitions.
In section VI we analyse the effects of decoherence during the computational
process at each node and show how they affect the results obtained in the 
preceding sections.
Finally, in section VII we summarise the main results of this work.

% ============================================================================
\section{Costs of Distributed Quantum Computation}

Let us start out by considering a generic scenario for distributed
quantum computation: a central processor $A$ and $n-1$ processors
labelled $B_i$ ($i=1,2,\ldots,n-1$) represent the nodes of a quantum
network. These nodes agree on performing a given computation which
consists of three steps: (1) {\it Precomputation}: In order to prepare
the initial state of all the nodes they exchange certain classical and
quantum information; (2) {\it Computation at each node}: Each of the
nodes performs a well defined operation locally, followed by a
measurement; (3) {\it Communication of the results}: The $B$ nodes
report the outcomes of their measurements to the central node by sending
$k$ bits of classical information. With this information,
the central node estimates the outcome of the complete computation. The
computation gives the correct result with certain probability, and
therefore it has to be repeated a number of times in order to achieve a
prescribed precision. 

We are interested in the cost of the computation in terms of the number of
uses of the processors, and the amount of
classical and quantum communication involved in the whole computation. 
Let us denote by
$P(n)$ the cost of the precomputation, i.e., the cost of establishing
the initial state for the $n$ nodes. In general, if one wants to create
an entangled state of all the nodes one has to send quantum information
through the channels. Besides, due to the presence of noise during
transmission and processing, the nodes will have to use either
purification or error correction methods, which will require in addition
some classical communication. Thus, the cost of precomputation 
will depend on the
costs of sending qubits and classical bits through the communication
channels that link the nodes. 
We denote by $Z$ the cost of running a quantum processor at each
node, and by $Y$ the cost of sending the outcomes of the measurement
from one node to the central node. Finally, we will denote by $R(n)$ the
number of times the computation has to be repeated in order to obtain a
prescribed precision. With these definitions, one can calculate the total
cost of the computation. We wish to analyze the advantages of using
entangled input states in the computation with respect to the case of
initial uncorrelated states. Thus, we consider the
following two scenarios:

{\em (1) Disentangled states}: If the initial state of the processors
is disentangled, no communication is required in the precomputation.
Therefore we take $P(n)=0$ \cite{note1} and obtain
\be
C_1(n)=R_1(n)[nZ+(n-1)Y].
\ee

{\em (2) Entangled states:} If the initial state of the processors
is entangled, communication is required in the precomputation. We have
\be
C_2(n)=R_2(n)[P_2(n)+nZ+(n-1)Y].
\ee

We can now evaluate the ratio between the cost $C_2(n)$ corresponding to 
entangled
inputs and the cost $C_1(n)$ for independent processors. We obtain
\begin{equation}
\frac{C_2(n)}{C_1(n)} = \frac{R_2(n)}{R_1(n)} 
\frac{P_2(n)+ (n-1) Y + nZ}{(n-1) 
Y + nZ}.
\label{cost}
\end{equation}
This ratio depends crucially on the ratio between the repetitions needed
in each case, as well as on the cost of the precomputation. Under ideal
conditions, the use of entangled states in general decreases the number of
repetitions required, i.e. $R_2(n)<R_1(n)$. The use of entangled states
will be cost efficient if what one gains in the number of repetitions
compensates what one loses in the precomputation. Thus, we expect that
there will be a certain $n_{min}$ such that if $n>n_{min}$, the use of
entangled states is cost efficient. On the other hand, in the non ideal
situation in which one has noise either in the quantum channel or
during the computation, $R_2(n)$ may increase with $n$ more rapidly than
$R_1(n)$ since entangled states are more prone to errors. Thus, we would
expect that for a specific task there is a maximum value $n_{max}$ such
that for $n<n_{max}$ the use of entangled states is cost efficient.
Therefore, if $n_{min} < n_{max}$ there will be an interval
$[n_{min},n_{max}]$ in which entangled states are more advantageous. In
the opposite case $n_{min} > n_{max}$, it will be more convenient to use
disentangled states. The values of $n_{min}$ and $n_{max}$
will depend on the specific computation and of the methods used to create
the entangled states. In the following we will concentrate on the specific
but relevant problem of phase estimation addressed in Ref. \cite{grover:97}. 
We will first analyze
the ideal noiseless case. Then we will consider 
the presence of noise in the quantum channel, and analyze two ways
of overcoming this noise using entanglement purification. Finally, we
will consider the effects of decoherence during the computation.

% ========================================================================
\section{The phase estimation problem}

As an illustrative and, in spite of its simplicity, important example,
consider a network of $n$ processors each performing computation $\cal
C$ defined as a small conditional phase shift on a qubit
\begin{eqnarray}
\ket{0} & \longrightarrow & \ket{0},\\
\ket{1} & \longrightarrow & e^{i\phi}\ket{1}.
\end{eqnarray}
Computation $\cal C$ is performed at each of the $n$ nodes ($A,\; B_i$).
Each run consists of a conditional phase shift and the subsequent
measurement protocol. Computation $\cal C$ is then {\em reset} after
each repetition: we assume that no extra-phase accumulation is allowed by
means of consecutive runs of the computation $\cal C$ on the same qubit
before the measurement is performed.

% ========================================================================
\subsection{Disentangled states}

Without inter-node entanglement, the best way to estimate $\phi$ is to
prepare each node in the initial state 
\begin{equation}
\ket{\Psi}_i = \frac{1}{\sqrt 2} (\ket{0}_i+e^{-i\phi_1}\ket{1}_i)\;.
\label{pulse}
\end{equation}
where $\phi_1$ is a given phase that can be adjusted from computation to
computation. Computation $\cal C$ is then applied, followed by a
Hadamard transformation $H$ given by $H = \frac{1} {\sqrt{2}} \left(
|0\rangle\langle 0| - |1\rangle\langle 1| + |0\rangle\langle 1| +
|1\rangle\langle 0|\right)$. The last step is the independent measurement of
each qubit in the computational basis. The result
of the measurement will be either $0$ or $1$ with probabilities $p_1$ and
$1-p_1$, respectively. Each of the $B_i$ nodes then transmits 
one classical bit, corresponding to the result of the measurement, 
to the central node $A$. This process is repeated $R_1$ times yielding
a binomial probability distribution. In this way one can estimate $\phi$ 
at node $A$ with precision
\begin{equation}
\epsilon_1 = \frac{\sqrt {\Delta p_1}}{|\frac{dp_1}{d \phi}| \sqrt R_1}\;,
\end{equation}
where $\Delta p_1=p_1(1-p_1)$ is the variance of the binomial distribution. 
In general, $\epsilon_1$ will depend on $\phi$ and $\phi_1$. As soon as
the first outcomes of the measurements are obtained (first repetitions) the
value of $\phi_1$ can be adjusted in order to minimise $\epsilon_1$ 
\cite{nist}.

% ========================================================================
\subsection{Entangled states}

Let us now assume that the initial state of the $n$ nodes is an
entangled state of $n$ qubits of the form
\begin{equation}
\ket{\Psi_{id}} = 
\frac{1}{\sqrt 2}(\ket{000\ldots 0}+e^{-in\phi_2} \ket{111\ldots 1})\;.
\label{ghz}
\end{equation}
Under ideal conditions, this state can be obtained as follows. The
central processor at the central node generates $n-1$ EPR pairs and sends
one member of each pair to the remaining nodes. An EPR pair shared
between node $A$ and $B_i$ is referred to as the $A-B_i$ pair.
In order to obtain the state in Eq. (\ref{ghz}) we pick up one
of the $n-1$ qubits at node $A$ and using it as a control qubit we apply
the quantum controlled-NOT operation
$|\epsilon_1\rangle|\epsilon_2\rangle \to
|\epsilon_1\rangle|\epsilon_1\oplus\epsilon_2\rangle$
($\epsilon_{1,2}=0,1$ and $\oplus$ denotes addition modulo 2) with the 
remaining $n-2$ target qubits at 
node A. Then we measure the $n-2$ targets in the computational basis.
At this stage we have already established an
entangled state of all $n$ nodes; in order to put it into the form 
(\ref{ghz}) we simply perform operation NOT ($\sigma_x$) at location
$B_k$ if the result of the measurement performed at $A$ on the qubit
belonging to the $A-B_k$ pair was $1$. Finally, all the nodes perform a 
phase shift transformation with angle $\phi_2$. 
In Figure 1 we have depicted a set-up for the simplest case, 
involving only three processors. Once EPR pairs of the form $\frac{1}{\sqrt{2}}
(\ket{0}_A\ket{0}_{B_i}+\ket{1}_A\ket{1}_{B_i})$ have been established
between nodes $AB_1$ and $AB_2$ (via channels represented in the
figure by a thick line networking the central node with the other two),
the central node $A$ performs a CNOT operation between the two qubits
stored in $A$. A measurement of the target bit in the computational basis
reduces the three-node composite state to either the state
$\frac{1}{\sqrt{2}} (\ket{0}_A\ket{0}_{B_1}\ket{0}_{B_2} +
\ket{1}_A\ket{1}_{B_1}\ket{1}_{B_2} )$ (outcome $0$) or to the state
$\frac{1}{\sqrt{2}} (\ket{0}_A \ket{0}_{B_1}\ket{1}_{B_2}+\ket{1}_A
\ket{1}_{B_1}\ket{0}_{B_2})$ (outcome $1$). In the latter case, node
$B_2$ has to invert the state of its qubit by means of a local
operation. Therefore, the pre-computation requires nodes $A$ and $B_2$
to exchange classical information, as illustrated in the figure by the
dotted line connecting those nodes.
% .........................................

Once a state close to the ideal state (\ref{ghz}) is prepared, 
at each node we run
computation $\cal C$ followed by the Hadamard transform. A measurement
on the computational basis is then performed at each node. Nodes $B_i$
report their outcomes to node $A$ by broadcasting one bit of information
and the overall parity of the reported bits and the outcome at node $A$
is calculated at $A$. This will give bit value $0$ or $1$ with
probabilities $p_2(\phi)$ and $1-p_2(\phi)$. The procedure is repeated
$R_2$ times and gives estimation of $\phi$ with precision
\begin{equation}
\epsilon_2 = \frac{\sqrt {\Delta p_2}}{|\frac{dp_2}{d \phi}| \sqrt R_2}\;,
\end{equation}
where $\Delta p_2=p_2(1-p_2)$. 
In general, $\epsilon_2$ will depend on $\phi$ and
$\phi_2$. As soon as the first outcomes of the measurements are obtained
(first repetitions) the value of $\phi_2$ is adjusted in order to minimise
$\epsilon_2$ \cite{nist}.

% ========================================================================
\subsection{Comparison}

In order to compare the two procedures, we impose that the precision required
is the same, i.e. $\epsilon_1=\epsilon_2\equiv \epsilon$. We obtain 
\be
\frac{R_2(n)}{R_1(n)} = \frac{\Delta p_2}{\Delta p_1} 
\left(\frac{dp_1}{d \phi}\right)^2 \left(\frac{dp_2}{d \phi}\right)^{-2}
\ee
where in this expression the phases $\phi_{1,2}$ are assumed to be 
chosen independently of each other
in order to minimize $\epsilon$. Once the values of $\Delta p_i$ and 
$dp_i/d\phi$ 
are known, one can substitute this expression in (\ref{cost}). The value of
$P_2(n)$ will depend on the purification procedures used in the generation
of the state (\ref{ghz}).

% ========================================================================
\section{Ideal Channels and Computations}

We consider first the simple situation in which no decoherence is present. 
In this case  
\bma
\bea
p_1 &=& \frac{1}{2}[1-\cos(\phi-\phi_1)],\\
p_2 &=& \frac{1}{2}[1-\cos(n\phi-n\phi_2)].
\eea
\ema
One obtains $\epsilon_1=1/(nR_1)^{1/2}$ and $\epsilon_2=1/(nR_2)^{1/2}$
independently of the values of
$\phi_{1,2}$. Therefore, we have $R_1(n)=1/(n\epsilon^2)$ and
$R_2(n)=1/(n\epsilon)^2$. On the other hand, the cost of the precomputation
using the procedure described above is simply $P_2(n)=(n-1)X+(n-2)Y$, where
$X$ is the cost of sending one qubit from the central node to any other node.
We finally obtain 
\begin{equation}
\frac{C_2}{C_1} = \frac{1}{n}\left(\frac{ (2 n - 3) Y + (n-1) X + nZ}{(n-1) Y + nZ}
\right).
\label{ratio}
\end{equation}
This expression implies that for $n$ larger than a certain value $n_{min}$, 
the global cost for computation with entangled states is smaller than the
one with independent states. It can be easily checked that
for $(2Y+X+Z)/(Y+Z)\gg 1$
\begin{equation}
n_{min} \approx \frac{2Y+X+Z}{Y+Z}.
\end{equation}
If the cost $Z$ is much smaller than $X$ and $Y$, and $Y$ is much smaller than
$X$, the threshold value
is given by the ratio of the costs of distributing entanglement 
and classical communication. Figure 2 illustrates this behaviour.

% ========================================================================
\section{Noisy Channels}

We have seen in the preceding sections that the use of entanglement is
cost efficient above a certain threshold in the number of nodes of a
quantum network. However, this result holds under the assumption that
the channels networking the nodes are ideal. In particular, this implies
that ideal entangled states $\ket{\Psi_{id}}$ can be
distributed among the $n$ nodes. In reality, this will never be the
case, and therefore one has to analyze what will happen for noisy
channels. While creating the state $\ket{\Psi_{id}}$ (using, for
instance, the protocol exemplified in Fig.\ 1), there will be errors.
The state will no longer be a pure state, but will rather be described
by a density operator $\rho\ne |\Psi_{id}\rangle\langle \Psi_{id}|$. The
closeness of this state to the ideal one is measured by the fidelity
$F_0=\langle\Psi_{id}|\rho|\Psi_{id}\rangle$. 
This means that the number of repetitions $R_2(n)$
required to perform the computation to a prescribed precision will
increase (since the probability distribution of obtaining the right
outcome becomes worse). On the other hand, one may use entanglement
purification in order to increase the value of $F_0$ and therefore to
reduce the number of required repetitions. However, this requires a
higher precomputation cost $P_2(n)$. In this Section we analyze this
problem for two different purification protocols, specially suited for
different situations. In order to focus on the role of 
noise along the channels, we will assume that all local operations (both
the ones required for the establishment of the entangled states and 
the ones involved in the computation at each node) are error free. 

We assume that using a two--qubit purification protocol and after $s$
steps one creates an entangled pair between nodes $A$ and $B_i$ with
fidelity $F_s\alt 1$. Let us denote by $P_1(s)$ the cost required to
create such a state. Once we have the entangled pairs we use the
method described above (see Fig. 1) to create the entangled state among the
$n$ nodes. Assuming that the pairs are in a Werner--like state, the fidelity 
of 
the $n$--qubit state will be $F_n\simeq F_s^{(n-1)}$. In order to estimate how
the results are affected by noise, we consider for simplicity 
\be
\rho = x_n \ket{\Psi_{id}}\bra{\Psi_{id}}+ \frac{1-x_n}{2^n} I,
\ee
where $F_n=x_n+(1-x_n)/2^n$. If we perform the computation with this state
instead of the ideal one we obtain
\be
p_2 = \frac{x_n}{2}[1-\cos(n\phi-n\phi_2)] + \frac{1-x_n}{2}.
\ee
Therefore, to estimate the parameter $\phi$ with precision $\epsilon$, the computation 
has to be run a number of times
\begin{equation}
R_2(n) = \frac{1}{n^2 \,\epsilon^2}\left(1+\frac{1-x_n^2}{x_n^2 \, 
\sin^2(n \phi-n\phi_2)}\right)
\end{equation}
which now depends on $\phi$ and $\phi_2$. We take the optimal choice of 
$\phi_2$, which gives 
\begin{equation}
R_2(n) = \frac{1}{n^2 \, \epsilon^2 \, x_n^2}\;,
\end{equation}
and obtain the cost ratio 
\begin{equation}
\label{rat}
\frac{C_2(n)}{C_1(n)}  = \frac{1}{n x_n^2} 
\left(\frac{ (n-1)P_0(s) + (2n-3) Y + nZ}{(n-1) Y + nZ}\right)\;.
\label{noise}
\end{equation}
Taking into account that $x_n\simeq F_n \simeq F_s^{(n-1)}$ we see that
for a sufficiently high value of $n$, this ratio will be as big as we
please. This implies the existence of a value $n_{max}$ such that for $n\agt
n_{max}$ there is no gain in using entangled states. In fact, it may
happen that there is no gain for any value of $n$. 
On the other hand, the value $n_{max}$ will depend on
$P_0(s)$ and $F_s$, i.e. on the specific purification procedure. Besides,
as before, there will be a value $n_{min}$ such that if $n\in[n_{min},n_{max}]$
then entangled states can be cost efficient. We will illustrate these
features for some specific purification protocols.

% =========================================================================
\subsection {Purification Scheme 1}

Let us assume that we can create pairs between
node $A$ (central node) and node $B_i$ ($i=1,...,n-1$) with fidelity
$F_0$. For simplicity we assume that they are in a Werner state. We
consider that local operations are perfect, and therefore one can use
the purification procedures of Refs. \cite{purif1,purif2}. One
can easily show that after $s$ successful purification steps the
fidelity will be 
\begin{equation}
F_s \le 1-(2/3)^s(1-F_0) \, \epsilon_0.
\label{epsi}
\end{equation}
In order to calculate $P_0(s)$ we note that to obtain one pair
of fidelity $F_s$ one uses up at least $2^{s-1}$ pairs, and performs
in each of them at least one operation. Therefore, $P_0(s)\ge 2^{s-1} U$,
where now $U$ denotes the joint cost per used pair. Upon substituting these
expressions in (\ref{rat}) we see that unless $U$ is much smaller than
the other costs or $F_0\simeq 1$, there is no gain at all by using entangled
states. 

% =========================================================================
\subsection {Purification Scheme 2}

We will evaluate now the cost associated with pre-computation when
the nodes are networked via photonic channels \cite{Innsbruck}. As in
the previous scheme, the maximum fidelity resulting in this case,
$F_s$, approaches $1$ with the number $s$ of purification steps
exponentially fast
\begin{equation}
F_s \simeq 1 - a^s(1-F_0),
\end{equation}
where $a<1$ is a constant. 
In this scheme the cost of establishing one entangled pair $P_0(s)=bs$ 
is proportional to the number of steps. Substituting these quantities in
Eq. (\ref{rat}) one can see the existence of an interval $[n_{min},n_{max}]$
in which the use of maximally entangled states is more convenient with
respect to uncorrelated ones.

This purification protocol is more efficient than the previous one since
$P_0$ does not scale exponentially with $s$. The reason for this scaling
law in the purification scheme 1 is that one discards one pair at each
purification step, which makes it very ineffective. Therefore, it is more
desirable to use a purification scheme giving a finite yield, as the
hashing or breeding methods \cite{DiVincenzo}. Moreover, the purification
protocols used here are just based on the ones developed for two qubits, 
which lead to the exponential dependence of $F$ on $n$. By using ideas
similar to the ones developed for quantum repeaters \cite{repeaters} it
may be possible to improve this exponential dependence.

% ============================================================================
\section{Noise in the Computational Process}

Let us now assume that at each node the computation itself is not
error-free but dephasing-type decoherence is present at a rate $g$,
namely a random phase is introduced in front of the component $\ket{1}$
of the qubit with probability $e^{-g t}$ at time $t$ (notice that if one
considers a quantum optical implementation, the results that we will
present in the following hold also in the presence of spontaneous
emission). When measured in the computational basis, the bit value $0$ 
will now be obtained with probabilities \cite{fs} 
\begin{equation}
 p_{1} = \frac{1}{2} [1 - \cos (\phi-\phi_1) e^{-g t_c}] 
\end{equation}
when dealing with independent processors and
\begin{equation}
 p_{2} = \frac{F}{2} [1 - \cos (n \phi-n\phi_2) e^{-n g t_c}] + \frac{1-F}{2}
\end{equation}
in the case of maximally entangled nodes of the form (\ref{ghz}). 
In the above
equations $t_c$ is the time required to perform computation ${\cal C}$
and will be regarded as a fixed parameter in the comparison of the two
schemes. To achieve resolution $\epsilon$ the computation $\cal C$
must be performed
\begin{equation}
R_1 = \frac{1}{n \epsilon^2} \frac{e^{2g t_c} - \cos^2 (\phi-\phi_1)}{\sin^2 (\phi-\phi_1)}
\end{equation}
times with independent processors and
\begin{equation}
R_2 = \frac{1}{n^2 \epsilon^2 F^2} \frac{e^{2n gt_c} - F^2 \cos^2 (n \phi-n\phi_2)}
{\sin^2 (n \phi-n\phi_2)}
\end{equation}
times when a maximally entangled input is distributed among the $n$
nodes. The relative cost of both procedures is no longer $\phi$
independent. Following the argument presented in the previous Section,
we select the controllable phases $\phi_{1,2}$ in both procedures in
such a way that the measured phase approaches $\pi/2$ when dealing with
uncorrelated inputs and $\pi/2n$ when one uses entangled states.
Therefore, we can write
\begin{equation}
R_1 = \frac{1}{n \epsilon^2} e^{2g t_c} 
\end{equation}
while
\begin{equation}
R_2 = \frac{1}{n^2 \epsilon^2 F^2} e^{2n gt_c}.
\end{equation}
The cost ratio is then given by 
\begin{equation}
\frac{C_2}{C_1}=  e^{2 g t_c(n-1)} \left(\frac{C_2}{C_1}\right)_{g=0}.
\end{equation}
As it can be seen from Figure 3, the effect of dephasing type
decoherence is negligible in the limit $g t_c \ll 1/n$. The net effect
of decoherence during the computational process is a further reduction
of the domain where the use of entanglement is cost efficient, being
$n_{max}(g \neq 0) < n_{max}(g=0)$.
% ===================================================================================
\section{Conclusions}

In summary, we have introduced the notion of a generic cost of physical
operations in distributed quantum computation. This parameter allows 
to quantify the efficiency of a quantum computation which can be run
separately on different quantum processors belonging to a quantum
network. Previous work \cite{grover:97,cleve} has shown that the use of
entangled states could be advantageous for certain computations.
However, it was not obvious that the cost of generating entanglement or
the inclusion of noise during the computational process, might not nullify
their advantage. We have shown that under certain circumstances a
quantum network supplied with a maximally entangled input yields a
smaller global cost than the one required when dealing with $n$
independent inputs, provided that $n$ belongs to a given interval $n\in
[n_{min},n_{max}]$. We have illustrated this for the case of phase
estimation. The values of $n_{min}$ and $n_{max}$ crucially depend on
the purification protocols used to establish the $n$-- processor
entangled states, as well as on the presence of decoherence processes 
during the computation. 

We are grateful to C. H. Bennett, L. Hardy, R. Jozsa, N. L\"{u}tkenhaus,
M. B. Plenio, S. Popescu and S. van Enk for helpful discussions. \par
This work was supported in part by the European TMR Research Networks
ERP-4061PL95-1412 and ERP-FMRXCT96066, Hewlett-Packard, The Royal
Society of London and Elsag-Bailey, a Finmeccanica Company. SFH
acknowledges support from DGICYT Project No. PB-95-0594 (Spain).

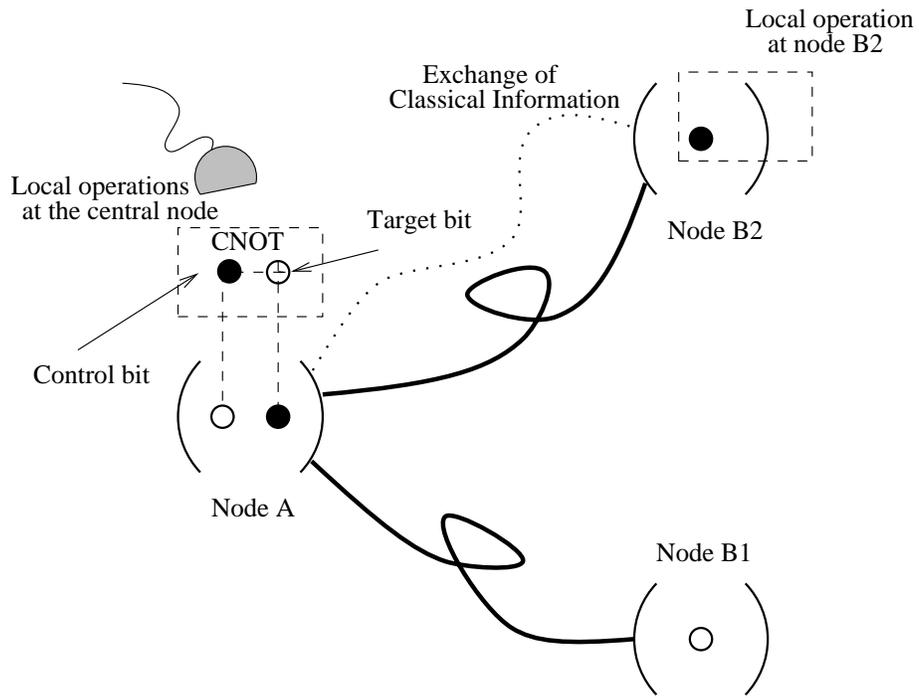
\begin{figure}[htb]
\begin{center}
\epsfxsize8.0cm
%\centerline{\epsfbox{grover.eps} }
\label{fig3}
\end{center}
\caption{\small Experimental set-up for performing a distributed quantum
computation between three nodes sharing an entangled state.}
\label{nodos}
\end{figure}
% ..........................................

% ...................................................
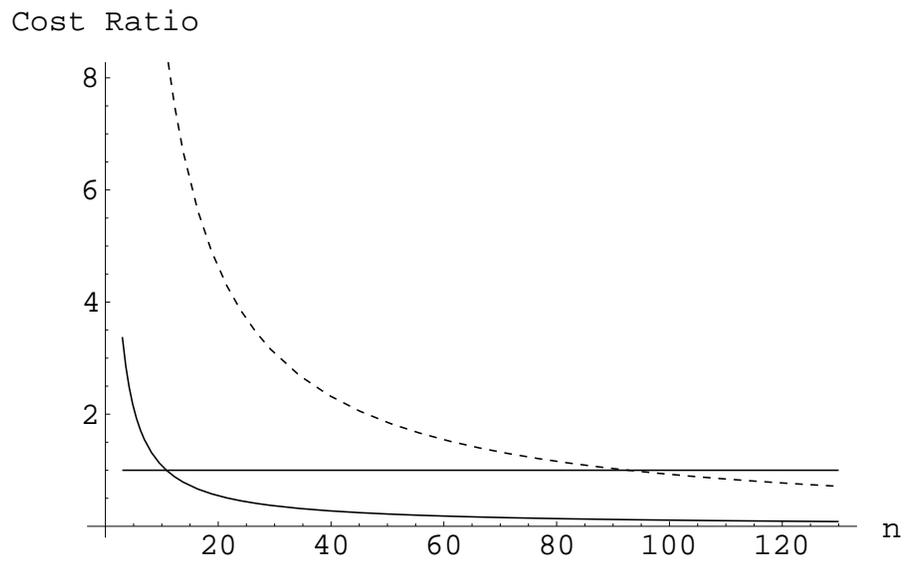
\begin{figure}[htb]
\begin{center}
\epsfxsize9.0cm
%\centerline{\epsfbox{qq.eps} }
\label{fig1}
\end{center}
\caption{\small Cost ratio $C_2/C_1$ as a function of the number of
nodes within the quantum network for two different sets of values of the
parameters $(X,Y,Z)=(100,10,1),(1000,10,1)$. Above a threshold $n_{th}
\approx X/Y$, the global cost of the computation is smaller when using a
maximally entangled input.}
\label{thres}
\end{figure}
% ...................................................

% .....................................................
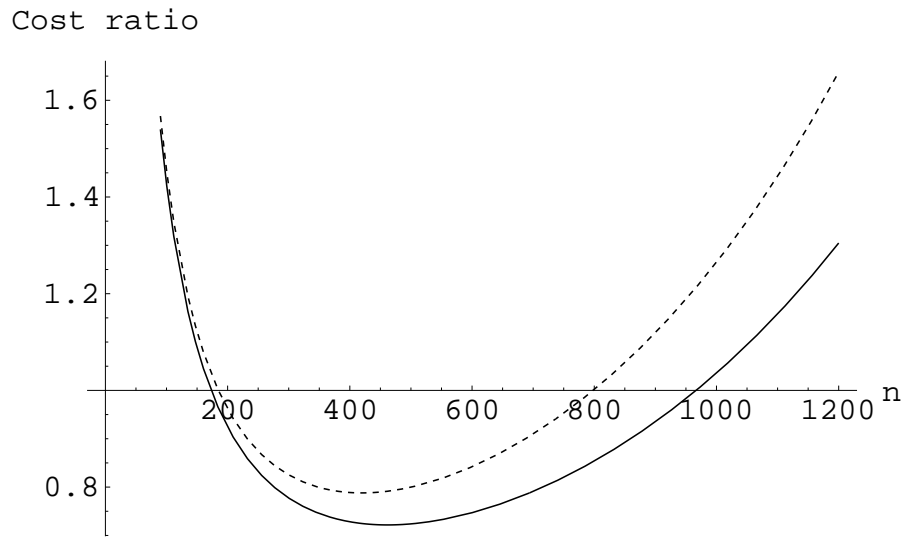
\begin{figure}[htb]
\begin{center}
\epsfxsize9.0cm
%\centerline{\epsfbox{gra2.eps} }
\label{figura3}
\end{center}
\caption{\small Cost ratio as a function of the number of nodes for a
phase estimation problem. The dotted line corresponds to the case when
dephasing type decoherence takes place during the computational process.
The net effect is that the domain where the use of entanglement is cost
efficient gets shrinked with respect to the case of error free
computation, represented by the solid line.}
\label{gra2}
\end{figure}
% .....................................................


\begin{thebibliography}{50}
\bibitem{note1}
We could consider the cost of creating the initial disentangling state at this
stage. In order not to overload the number of parameter and concentrate in
the differences caused by the use of entanglement we will omit this cost.
We will also omit here the cost of local operations, although they could
be easily included. The conclusions of our analysis do not change if one
includes these costs.
\end{thebibliography}
\end{document}